\documentclass[prb,preprint,showpacs,superscriptaddress,floatfix]{revtex4}
\usepackage{graphicx} 
\begin{document}
\draft
\title{Electric field in hard superconductors with arbitrary cross section and general critical current law}
\author{A. Bad\'{\i}a\,--\,Maj\'os}
\email[Electronic address: ]{anabadia@unizar.es}
\affiliation{Departamento de F\'{\i}sica de la Materia Condensada--I.C.M.A., C.P.S.U.Z., Mar\'{\i}a de Luna 1, E-50018 Zaragoza, Spain}
\author{C. L\'opez}
\affiliation{Departamento de Matem\'aticas, \\ Universidad de Alcal\'a de Henares, E-28871 Alcal\'a de Henares, Spain}
\date{\today}
%
%
\begin{abstract}

The induced electric field $\vec{E}(\vec{x})$ during magnetic flux entry in superconductors with arbitrary cross section $\Omega$ and general critical current law, has been evaluated by integration along the vortex penetration paths. Nontrivial flux motion streamlines are obtained from a variational statement of the critical state, which takes the form of an optimization problem on the finite element discretization of $\Omega$. The generality of the theory allows to deal with physical conditions not considered before. In particular, it is shown that the boundary condition to be used for determining $\vec{E}$ is the knowledge of the locus $E=0$ within the sample. This is emphasized for anisotropic materials in which the electric field is not parallel to the surface.
Both numerical and analytical evaluations are presented for homogeneous materials with different geometries: convex and concave contours, samples with holes, variable curvature contours, and for anisotropic samples. In the isotropic case, discontinuities in the electric current paths are shown to be related to changing curvature of the sample's surface. Anisotropic samples display the same kind of discontinuities, even for constant surface curvature.

\end{abstract}
\vspace{10mm}
\pacs{74.25.Sv, 74.25.Ha, 41.20.Gz, 02.30.Xx}
\maketitle
%
%
\section{Introduction}
Type II superconductors in the Critical State (CS) hold the key property of allowing non dissipative volume electric currents. These can be directly fed through contact pads, or induced by a penetrating external magnetic field. Bean's one dimensional model for the CS\cite{bean} is a very simple and successful tool for determining the metastable states of a magnetized sample. Such states are described by the spatial distribution of current densities and magnetic field. The operational advantage of this model is its sharp simplification against the use of the complete Maxwell equations set, plus an accompanying constitutive law $\vec{E}(\vec{J})$ for the material. However, the calculation of transitient electric fields during the magnetization process has attracted much interest. In particular, the knowledge of $\vec{E}$ allows to evaluate the local power dissipation $\vec{J}\cdot\vec{E}$, which is a basic quantity if one considers the thermal stability of the superconducting state.

Although $\vec{E}$ is not an explicit variable (it is customary bypassed) in the critical state theory of hard superconductors,\cite{bean} a quasistationary model for the hysteresis losses allows to estimate it for such case, i.e.: $E\simeq \delta\Phi/\delta t/\delta\ell$. Hence, $E$ is locally given by the amount of magnetic flux which has crossed the unit length per unit time. This idea was already used by Swartz and Bean\cite{swartz} in order to produce a pioneering model for magnetothermal instabilities. More recently, it was brilliantly applied for predicting the observed flux jumps in the magnetization of high $T_c$ superconductors.\cite{mueller} However, as a shortcoming of the method, $\delta\ell$ must be taken perpendicular to the magnetic flux paths, which are not obvious, unless for very simple geometries (infinite slabs and circular cylinders). In addition, one should consider the general case, in which $\vec{J}$ and $\vec{E}$ are not parallel (anisotropic material), and the fully vectorial information is required. 

Remarkably, as demonstrated by Brandt and co-workers\cite{brandt,schuster} for the case of superconductors with rectangular (and related) cross sections, the exact details of the function $\vec{E}(\vec{x})$ strongly depend on the sample's geometry. This was shown by analytical critical state calculations, as well as numerically evaluated, for the power law $E\sim J^{n}$, and experimentally tested. On the other hand, the relevance of the local behavior of  $\vec{E}(\vec{x})$ has been recently shown\cite{jooss} in thin film experiments. Two order of magnitude variations of $E$ have been reported, indicating that averaged values may be quite meaningless.

In this work, we propose a numerical method for the theoretical evaluation of $\vec{E}(\vec{x})$, based on general critical state principles\cite{badiaprl} and well suited to problems with arbitrary geometry. Since the pioneering finite element oriented modelling of superconductors proposed in Ref.\onlinecite{bossavit}, a detailed analysis of transitient electric fields has been lacking. Furthermore, within our theory, very general critical current restrictions may be considered. In fact, the material law for the superconductor is given in the form $\vec{J}\in\Delta$, with $\Delta$ some bounded set around the origin. 

For each time step, the  data for calculating $\vec E$ are the corresponding magnetic field distributions before and after the changes, as determined by our variational model.  Then, the flux penetration paths can be derived and, in two dimensional problems, a simple line integration allows to obtain the electric field at every point of the sample where the profile has changed ($\vec E$ vanishes otherwise). The method maintains the advantages of the variational principle. It has a simple formulation for general arbitrary geometries, it allows to consider anisotropic models, and it has a convenient numerical implementation on the finite element discretization. 

By application of the previous ideas to several cross sections, we report on the complex $\vec{E}(\vec{x})$ structure under various circumstances. In particular, we focus on discontinuity lines, originated either by corners, inhomogeneities or just by changing curvature in the case of smooth sample boundaries. Finally, we report on some pecularities for the case that flux paths are not perpendicular to the electric current flow. This property is outstanding for anisotropic material in which the current flow is forced away from the crystal principal axes, as opposite to the configurations considered before.\cite{schuster2}
%
%
\section{Theory}
\subsection{Mathematical Statement}
Recall that the variational interpretation of the longitudinal critical state for a long superconductor with arbitrary cross-section $\Omega$ reads\cite{badiaprl,badiah}
\begin{equation}
\label{eqnvarpri}
{\tt minimize }\quad\frac{1}{2}\int_{\Omega}\! | H_{\rm n+1} -
H_{\rm n} |^{2}\;d\Omega\quad {\rm for} \quad\vec{u}\in\Delta_{\perp} \; ,
\end{equation}
with $H_{\rm n+1}$ the unknown magnetic field at the time layer $({\rm n+1})\delta t$, $\vec{u}={\bf grad} H_{\rm n+1}$, and $\Delta_{\perp}$ a $90^{o}$ rotation of the restriction set $\Delta$ for the current density. In fact, if the superconductor is parallel to the Z-axis, $\vec{u}=(-J_{y},J_{x})/J_{c}$. Notice that the typical range in which equilibrium magnetization and surface barrier effects can be neglected is assumed, i.e.: $\vec{B}\simeq\mu_{0}\vec{H}$. 

The solution of Eq.(\ref{eqnvarpri}) may be investigated by the Optimal Control theory, as shown in the above mentioned articles. On following this formalism, we define a Hamiltonian density
\begin{equation}
{\cal H}({H}_{n+1}, {\vec u}, {\vec p}, \vec{x})\equiv{\vec p}\cdot {\vec
u}-{{1}\over {2}}|{H}_{n+1}-{H}_n|^2 \; ,
\end{equation}
which contains the Lagrangian density to be minimized, as well as the associated auxiliary momentum $\vec{p}$. Then, the solution of the problem (superindexed by *) verifies the canonical system
\begin{equation}
\label{eqnham1}
{\bf grad}\;{H}_{n+1}^*=  \frac{\partial{\cal H}}
{\partial \vec{p}^{\,*}} = {\vec u}^{\,*} \; ,
\end{equation}
\begin{equation}
\label{eqnham2}
{{\rm div}\;{\vec{p}}^{\,*}} = - \frac{\partial {\cal H}}{\partial
{H}_{n+1}^*}=
{H}_{n+1}^*-{H}_n(\vec{x})
 \; ,
\end{equation}
with ${\vec u}^{\,*}$ determined by the algebraic condition of maximality
\begin{equation}
\label{eqalg}
{\cal H} ({H}, {\vec u}^{\,*}, {\vec p}, \vec{x}) \geq {\cal H} ( {H}, {\vec u}, {\vec p}, \vec{x})
\qquad \forall \;{\vec u} \in \Delta_{\perp} \; .
\end{equation}
Owing to the linear dependence of ${\cal H}$ on ${\vec u}$ (the Lagrangian density does not explicitely depend on this variable), the solution may be characterized by a {\em maximum projection} rule
\begin{equation}
\label{eqmaxpro}
\vec{p}\cdot{\vec u}^{\,*} \geq
\vec{p}\cdot{\vec u}\qquad \forall \;{\vec u} \in \Delta_{\perp}
\; ,
\end{equation}
and the critical state is given by a control vector lying on the boundary of the allowed control set ($\vec{u}^{\,*}\in\partial_{0}\Delta_{\perp}$). Recall that the generalized concept of the boundary $\partial_{0}\Delta_{\perp} = \{\partial\Delta_{\perp}\}\bigcup\,\{\vec{0}\}$ must be used.\cite{badiah} Recall also that, in the case that $\Delta_{\perp}$ is a circle this leads to the standard critical state expression $|\vec{J}|=J_{c},0$.

We want to emphasize that, from the mathematical point of view, Eq.(\ref{eqmaxpro}) is valid for any shape of the region $\Delta$. Physically, this means that our theory can host a variety of CS models, as it was emphasized in Ref.\onlinecite{badiajltp}. In previous work, this generality has been exploited for simulating the quasistatic magnetization properties of hard superconductors under various conditions. Here, we show that this may also be applied to the transitient electric fields.

\subsection{Physical interpretation: electric field}

Below, we show that the mathematical variable $\vec{p}$ (Lagrange multiplier of the theory) is physically meaningful if Eq.(\ref{eqnham2}) is interpreted as a discretized {\em continuity equation} for the magnetic field. In fact, in the absence of either sources or sinks one may consider the {\em transport equation}
\begin{equation}
\frac{\partial H}{\partial t}+{\rm div}\vec{J}_{H}=0  \; .
\end{equation}
Then, $\vec{p}$ is simply related to the {\em magnetic field current density} by
\begin{equation}
\vec{p}\equiv -\vec{J}_{H}\,\delta t \;\; .
\end{equation}
Let us further exploit this equivalence, which will be the basis for obtaining the transitient electric field when the critical state is perturbed by the external excitation. First, notice that Eq.(\ref{eqnham2}) may be written as (superindices will be dropped hereafter, because the optimal solution is assumed)
\begin{equation}
\label{eqdivp}
{\rm div}\,\vec{p}={H}_{n+1}- {H}_{n}={\bf grad}\,p\,\cdot\hat{p}
+p\;{\rm div}\hat{p}  \; .
\end{equation}
Now, define the vector field $\vec{e}\equiv\vec{p}\times\hat{z}$. It is apparent that one has
\begin{equation}
\label{eqrote}
{\bf rot}\,\vec{e}\cdot\hat{z}=-({H}_{n+1}- {H}_{n})=-{\bf grad}\,e\,\cdot\hat{p}
-e\;{\rm div}\,\hat{p}  \; .
\end{equation}
This equation is the main result of our work, as it establishes the connection between the variational problem solution [$H_{\rm n+1}(\vec{x})$ and $\hat{p}(\vec{x})$] and the physical variable $\vec{e}$. In fact, notice that $\vec{e}$ is nothing but the induced electric field in the evolution $H_{\rm n}\to H_{\rm n+1}$ if appropriate units are used, i.e.: Eq.(\ref{eqrote}) represents a convenient discretized form of the induction law $\vec{\nabla}\times\vec{E}=-\dot{\vec{B}}$. 

We call the reader's attention that the main advantage of using Eq.(\ref{eqrote}) relies on the fact that it contains the full vectorial information on the vortex trajectories (given by $\hat{p}$), and this allows to derive $\vec{E}$ for nontrivial cases, within the CS framework. Previous studies were restricted to straight line flux paths (square and circular symmetries).

As regards the boundary conditions to be used, the constraint that $\vec{E}$ is parallel to the sample's surface is not a sufficient condition for determining this vector field, as it was suggested in Ref.\onlinecite{brandt}. On the one side, this condition is only valid for the isotropic case, in which $\vec{E}\parallel~\!\!\vec{J}$. Furthermore, for an arbitrary shape of the restriction region $\Delta$, even the direction of $\vec{E}$ is unknown along the surface and has to be calculated {\em a posteriori}. Generally speaking, $\vec{E}$ must be determined by the {\em specification of the locus} $e=p=0$. In physical terms, this locus is defined by two kinds of lines. First, we have $e=0$ for the free boundary reached by the penetrating front ($H_{\rm n+1}=H_{\rm n}$). In addition, $e$ must vanish at the points where the vector field $\vec{e}$ displays sharp bends. Macroscopically, this follows from the continuity condition for the tangential component of $\vec{E}$ (provided $\dot{B}$ is finite, as it should be in the absence of flux cutting phenomena!). 

As soon as the critical profile $H_{\rm n+1}(\vec{x})$ has been obtained, one may cast Eq.(\ref{eqrote}) in the form
\begin{equation}
\label{eqelectric}
\delta e=\left (
H_{\rm n+1}-H_{\rm n}-e\,{\rm div}\hat{p}
\right )
\delta s  \; ,
\end{equation}
with $s$ the arc length measured along the streamlines of $\hat{p}$. Notice that $\hat{p}$ (flux penetration paths) may be obtained by combination of Eqs.(\ref{eqnham1}) and (\ref{eqmaxpro}) in the form $\hat{p}(\vec{u})$. Thus, an electric field map may be obtained just by integration with starting points at the lines $e=0$. 

In order to obtain an explicit analytical form of Eq.(\ref{eqelectric}), one must select a given restriction for the current density. For instance, within the isotropic model [$|\vec{J}|\leq J_{c}$ (i.e.: $\Delta$ is a circle)] Eqs.(\ref{eqnham1}) and (\ref{eqmaxpro}) lead to ${\bf grad}H_{\rm n+1}= \vec{u}=\hat{p}\,$. Insofar as in this case $\vec{e}=e\vec{J}/J_{c}$, and as the current density streamlines are strightforwardly determined by the critical state solution, one is just led to solve for the {\em scalar field} $e(\vec{x})$, and one equation suffices. We get
\begin{equation}
\label{eqinte}
\delta e=\left(
H_{\rm n+1}-H_{\rm n}-e\,\nabla^{2}H_{\rm n+1}
\right)
\delta s  \; ,
\end{equation}
which is a quite simple expression in terms of the critical state profiles. Other selections of $\Delta$, however, lead to cumbersome expresions and are better treated just at the numerical level, keeping Eq.(\ref{eqelectric}) as the basis. As an example, elliptic anisotropy is treated later in this work.
%
%
\section{Application: isotropic cases}
Below, we give the results of the previous procedure for isotropic samples with cross sections in the form of (i) a square with circular holes, (ii) a star with sharp concave corners, and (iii) an ellipse. These examples have been investigated numerically, by application of Eqs.(\ref{eqnvarpri}) and (\ref{eqinte}). The actual numerical method relies on the finite element discretization of our variational statement, which was described elsewhere.\cite{badiah} Basically, Eq.(\ref{eqnvarpri}) becomes a matrix quadratic optimization problem, when the magnetic field is expanded in terms of nodal mesh functions. This can be solved by a number of computational mathematics algorithms.\cite{lancelot} In addition, comparison with simple analytical criteria has been done when possible.
\subsection{Square cross section with circular hole}
Let us first consider an increasing external field applied parallel along the axis of a squared section cylinder (edge lenght $2a$, and origin of coordinates  at the center) with a cylindrical hole (radius $R$ and center $(0,y_c)$). As one can see in Fig.\ref{fig1}, the initial fronts of penetration are parallel to the edges, determining successive smaller and smaller squares, $x= \pm x_o$, $y= \pm y_o$, and the usual {\em sand pile} profile. The intersection of two perpendicular fronts produces a line of discontinuity for the current density $x= \pm y$, and a vanishing electric field.\cite{brandt} Recall that field contours and current streamlines are the same thing for the long specimen geometry.
 
As soon as the flux fronts reach the hole $y_o = y_c + R$, it is filled with a uniform value of the penetrating field. Afterwards, the straight fronts associated to the edges are accompanied by a circular front $x^2 + (y-y_c)^2= r_o^2$ emanating from the hole. The intersection of the planar fronts from the edges and the circular one determine parabolic lines of discontinuity for the vector $\vec{J}$, given by $x^2 = 4 R (y_c + R - y)$ (vertical parabola) and $(a-y_c-R)^2 + (y-y_c)^2 = 2(a-y_c+R)(a-y_c-R\pm x)$ (horizontal parabolae) as it is apparent in Fig.\ref{fig1}.

The electric field can be obtained from Eq.(\ref{eqinte}), by integration with starting point at the lines $\vec{e} = 0$, towards the boundary. For illustration, Fig.\ref{fig2} displays a number of flux penetration streamlines obtained with our numerical method for a given field step. In addition, a 3D plot of $e$ is given, showing that one gets the expected behavior near the edges (compare to Ref.\onlinecite{brandt}) as well as a considerable increase around the hole. Notice that a nondisplayed high flow of magnetic field is concentrated around the line $x=0$, $a \geq y \geq y_c+ R$, which connects the hole with the nearest edge, because it is the line by which the hole is being refilled. Actually, for the ideal case of a perfect circular hole, $E$ diverges as $E\simeq \delta\Phi/\delta t/\delta\ell$ and $\delta\ell\to 0$.

Some additional remarks, concerning the physics of flux penetration in samples with holes, may be done by analyzing the case of a lattice of circular holes. Fig.\ref{fig3} shows the current density contours and associated $e=0$ line structure for a $3\times 3$ square lattice. It is apparent that, soon after the vortices have reached the outer row of holes, a physical boundary is established (continuous line in the figure) where the flux velocity becomes zero. Subsequent penetration of flux towards the sample's core only takes place across the intersections of this boundary and the holes, which behave as a set of point sources with interfering circular fronts. Further flux motion barriers are estalished at the intersections between vortex trajectories emanating from the holes (straigtht dashed lines in Fig.\ref{fig3}).
 
\subsection{Tetracuspid shaped cross section}
On taking the four points $(\pm R , \pm R)$, and drawing the four quarter circumferences of radius $R$ closest to the origin we obtain a {\em tetracuspid} like region with sharp spikes (see Fig.\ref{fig4}). Being it the cross section of a long superconducting sample, we may consider the problem of an increasing applied magnetic field and determine the penetrating profile as well as the generated transitient electric field inside the sample. Fig.\ref{fig4} shows the penetrating magnetic field contours, as well as the electric field corresponding to the transition from the displayed to the subsequent critical state.
 
The main property of this example is that $\vec{e}$ is a nonlinear function of space and, as a consequence, the induced electric charge density during the transition is not piecewise constant as reported in Ref.\onlinecite{brandt} for the rectangular geometry. In fact, on using a cylindrical coordinate system, with the origin at the center of a circumference (f.i.: $(-R,-R)$) $q$ may be analytically evaluated (just one quarter of the problem needs to be considered by virtue of the symmetry)
\begin{eqnarray}
q=\left\{
\begin{array}{cr}
\epsilon_{0}\,\dot{B}{R^{2}\sin{\theta}}/{r^{2}\cos^{3}{\theta}}\quad, & \quad 0<\theta<\theta^{*}\\
0\quad,  & \quad \theta^{*}<\theta<\pi/2-\theta^{*}\\
-\epsilon_{0}\,\dot{B}{R^{2}\cos{\theta}}/{r^{2}\sin^{3}{\theta}}\quad,  & \quad \pi/2-\theta^{*}<\theta<\pi/2
\end{array}
\right.
\end{eqnarray}
Above, the critical angle $\theta^{*}$ refers to the intersection between the penetrating free boundary and the {\em current bending} discontinuity lines (see Fig.\ref{fig4}). In the full penetration regime, we get $\theta^{*}=\pi/4$ and a piecewise continuous behavior for the induced charge density, with jumps from positive to negative values at $\theta^{*}$.

\subsection{Elliptical cross section}
\label{secellip}
The long superconductor with elliptical cross-section was already considered by Campbell \& Evetts\cite{campbellevetts} as a model system for the critical state. It was illustrated that when the flux front reaches the nearest center of curvature of any part of the surface, further advance in that direction is halted, with a resulting cusp in the subsequent field distribution. Here, we show that our variational statement of the critical state reproduces this behavior and allows an easy computation of penetration fronts, which have a by no means trivial analytical evaluation, unless some approximation is used. In this sense, we should mention that, based on the assumption that flux penetration has the shape of an ellipse, a very accurate analytical solution was proposed.\cite{mikitik}  Additionally, we can give a detailed study of the arising electric field, as well as the $e=0$ line structure. 
Fig.\ref{fig5} shows the evolution of penetrating flux. On using $a$ and $b$ for the ellipse semiaxes, the closest curvature centers are located at $(\pm a\mp b^{2}/a,0)$. This points have been marked, and one can verify that the current streamline structure is modified as expected when flux goes through.

The $e=0$ structure is also peculiar to this example. As one can see in Fig.\ref{fig5}, the flux penetration streamlines are concentrated around the closest curvature centers, display an increasing angle along the straight $e=0$ segment and, eventually, reach the free boundary line at an angle of $\pi /2$. 

We want to notice that this example provides the explanation of a very general fact. The $E=0$ lines related to electrical current path discontinuities will arise for any contour with changing curvature. Obviously, this will be always the case, unles for circular cylinders. Furthermore, such discontinuity  lines are always confined within the sample. The case of corners must be considered just an idealization for which the maximum curvature centers have reached the boundary itself.
%
%
\section{Application: anisotropic behavior}
As discussed in the theoretical section of the paper, the proposed method for obtaining the mean electric field distribution from the variational CS theory is quite general, and includes arbitrary cross sections as well as anisotropic material, among other features. In this respect, we recall that our theory was already applied to anisotropic samples,\cite{badiajap} for the analysis of frequency mixing phenomena in rotation experiments, but restricted to the infinite slab geometry.

Here, as a final application, a circular cross section with anisotropic material law $J_x^2/\gamma^2~+~J_y^2~\leq~J_c^2$ ($\Delta$ is an ellipse) is presented (see Fig.\ref{fig6}). As it was discussed in Ref.\onlinecite{badiajap} this law within our variational interpretation is fully equivalent to a perturbation theory of isotropic media, which satisfies the basic requirement $\nabla_{\!\vec{E}}\times\vec{J}=0$, i.e.: the $\vec{J}({\vec{E}})$ relation is free from local loops.\cite{mayergoyz}

The first observation in our simulations is that the vortex penetration paths within the sample (along $\hat{p}$ directions) are no longer perpendicular either to the surface or to the electric current streamlines. In particular, as shown in the figure, this means that the electric field generated at the surface has both tangential and normal components, showing a broader generality of the method against previous studies. In more detail, if one applies the maximum projection rule [Eq.(\ref{eqmaxpro})] of our variational principle, flux motion lines are given by the vector field
\begin{equation}
\label{eqpaniso}
\hat{p} = \frac{(\gamma ^2 \partial H/\partial x, \partial H/\partial y)}{\sqrt{\gamma ^4 (\partial H/\partial x)^2 + (\partial H/\partial y)^2}}
= \frac{(-\gamma ^2 J_y, J_x)}{\sqrt{\gamma ^4 {J_y}^2 +{J_x}^2 }} \; .
\end{equation}
This equation allows to evaluate the angles between the physical quantities involved. For instance, the angle between $\vec{E}$ and $\vec{J}$ is determined by
\begin{equation}
\label{eqanglej}
\cos{\alpha}= \frac{J_{x}^{2}+\gamma^{2}J_{y}^{2}}
{J\sqrt{(J_{x}^{2}+\gamma^{4}J_{y}^{2})}}  \; .
\end{equation}
In this particular example, a change of coordinates allows to transform the anisotropic law into an isotropic one, mapping the circular section into an elliptic section, and obtaining a correspondence between apparance of cuspidal points for the current flow. Thus, on choosing a given value for $\gamma$, we can straightforwardly compare the results of this section to the example in Sec. \ref{secellip} for $b/a=\gamma$. This property was already exploited in Ref.\onlinecite{mikitik}, but restricted to the examination of the current streamlines and flux fronts. In our case, we have calculated the cuspidal points for the anisotropic system, which are plotted in Fig.\ref{fig6} just for validation of the numerical method. More relevant from the point of view of this work is that anisotropic material laws and variable surface curvature geometry have equivalent effects on the electric field evolution. In particular we show the appearance of the inner points where the $E=0$ lines associated to sharp bends in the current flow arise. However, contrary to the isotropic case, the vortex trajectories are no longer perpendicular to the current streamlines and $\vec{E}$ is no longer parallel to $\vec{J}$ [see Eq.(\ref{eqanglej}) for the actual angle between these quantities].  
%
%
\section{Summary}
A numerical method is presented which allows the computation of induced local electric fields during magnetic flux penetration (or exit) in hard superconductors. The theory is based upon a variational interpretation of the critical state, from which one can derive the flux penetration streamlines (vortex trajectories) in nontrivial cases. Integration along these lines, starting at the $E=0$ points, and ending at the sample's surface, is used for calculating the electric field. The generality of the theory allows to investigate the physical properties related to an arbitrary sample's cross section, as well as the implications of considering a wealth of critical state models. In fact, the critical current law is formulated as the very general restriction $\vec{J}\in\Delta$ for the current density.

The application of our algorithm has allowed a quantitative study of (i) the electric field in the vicinity of a hole near the edge of a square sample, (ii) the simulation of nested flux barriers in the magnetization of samples with a square lattice of holes, (iii) the properties of $\vec{E}$ for samples with sharp concave spikes, (iv) the discontinuity lines for the current density, as related to changes in the surface curvature, and (v) electromotive forces arising from vortex penetration non perpendicular to the electrical current streamlines (anisotropic regime).

The wide application range of the theory allows the expeditious incorporation of new physical phenomena as spatial inhomogeneities in the pinning force, or flux cutting interactions for non parallel vortex lattices (just by selecting $\Delta$ properly).
\section*{Acknowldegements}
This work has been supported by the Research Groups Budget of the local Government (region of Arag\'on) and the Research Program of the University of Zaragoza. Financial support from Spanish CICYT (Project No. BFM2003-02532) is also acknowledged.

The authors are indebted to Leonid Prigozhin for helpful discussions along several stages of this work.
\newpage
\centerline{\large\sc REFERENCES}
\newpage
%
\centerline{\large\sc FIGURE CAPTIONS}
\begin{figure}[h]
\caption{\label{fig1}Current density streamlines in the critical state for a square superconductor with a circular hole close to the upper edge. Four stages of the initial magnetization process are depicted. The diagonal $e=0$ lines have been marked in the upper left picture and dashed parabolic $e=0$ lines around the hole for the complete set.}
\end{figure}
\begin{figure}[h]
\caption{\label{fig2}Flux penetration streamlines (left half) and induced electric field modulus (right half) in the magnetization process of a square superconductor with a circular hole close to one edge. Parabolic and diagonal $e=0$ lines have been marked.}
\end{figure}
\begin{figure}[!]
\caption{\label{fig3}Current density streamlines in the critical state of a square superconductor with a ($3\times 3$) square lattice of holes. The $e=0$ structure has been emphasized by dark lines. Continuous style is used for the flux penetration barrier and dashed for connections with inner boundaries.}
\end{figure}
\vspace{-3cm}
\begin{figure}[!]
\caption{\label{fig4}Upper left: Current density streamlines in the critical state for a superconductor with tetracuspid like cross section; right: induced electric field modulus contours for increasing magnetic field. Lower left: flux penetration streamlines; right: 3D plot of the induced electric field modulus, obtained by integration along the streamlines. Only one quarter is depicted for clarity. }
\end{figure}
\begin{figure}[!]
\caption{\label{fig5}Flux penetration in a superconductor with elliptical cross section in the critical state. Upper: current density streamlines previous (left) and subsequent (right) to the contact of the flux front with the maximum curvature centers. Lower left: detail of the flux penetration paths around a maximum curvature point. This point and $e=0$ lines have been highlighted. Lower right: 3D plot of the transitient electric field modulus for increasing magnetic field. For clarity, one half of the problem is shown.}
\end{figure}
\begin{figure}[!] 
\caption{\label{fig6}Flux penetration in a superconductor with circular cross section and anisotropic critical current. Upper left: current density streamlines subsequent to the contact of the flux front with the cuspidal points (see text). Upper right: critical current distribution. Lower left: flux penetration paths in the neighbourhood of the cuspidal point. Lower right: 3D plot of the transitient electric field modulus. The arrows indicate the direction of $\vec{E}$ on the sample surface.}
\end{figure}
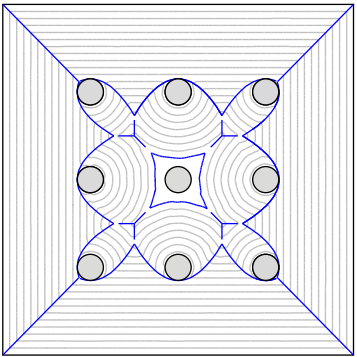
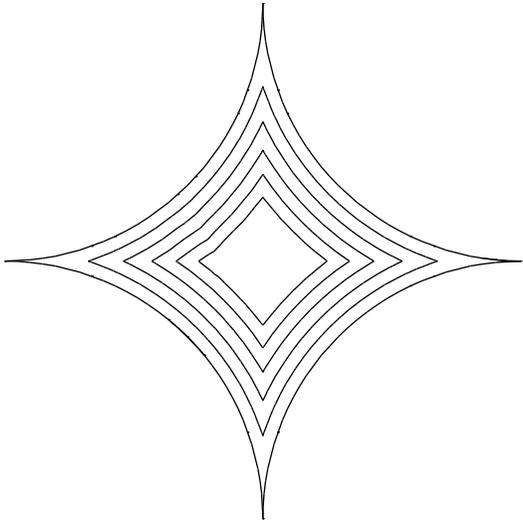
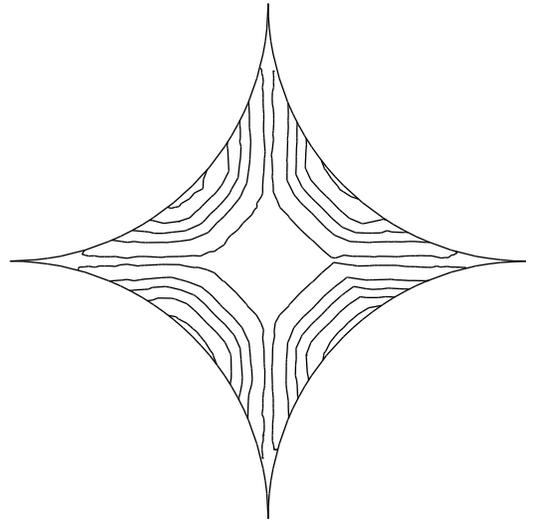
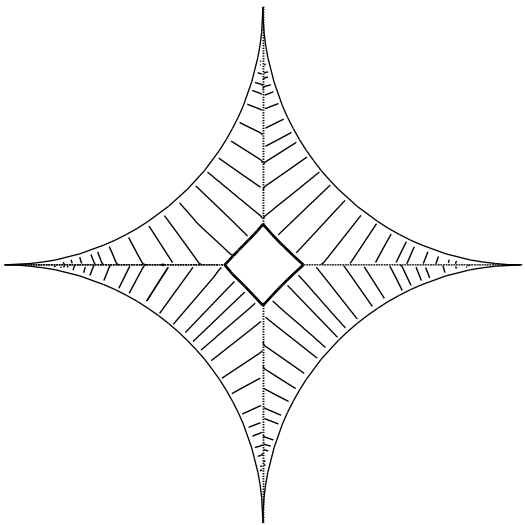
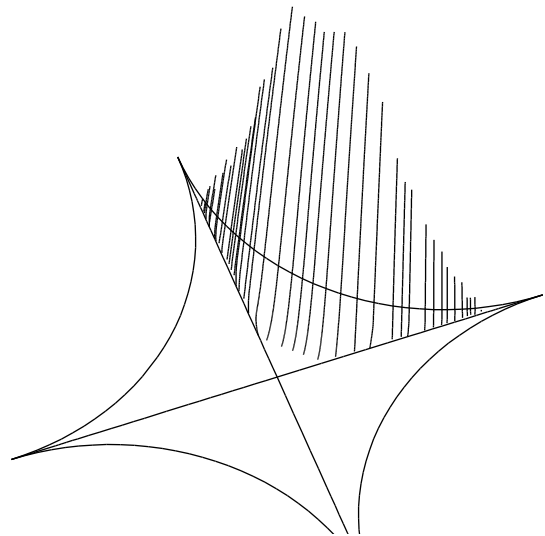
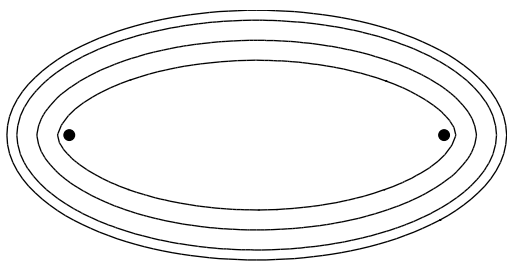
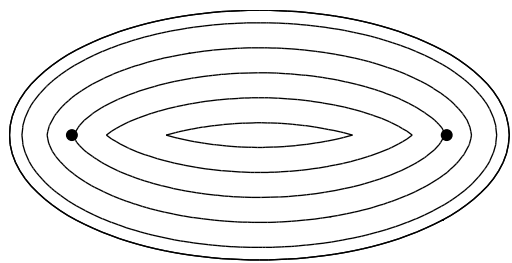
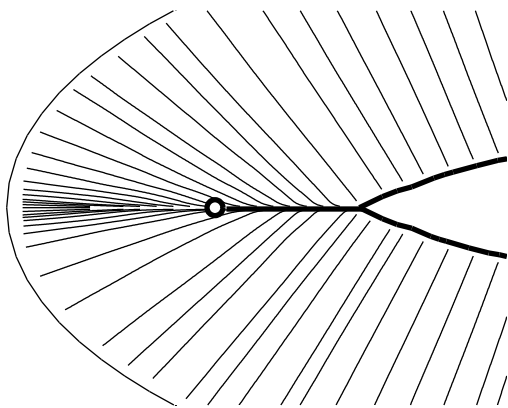
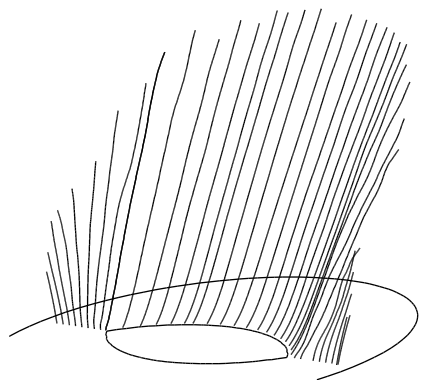
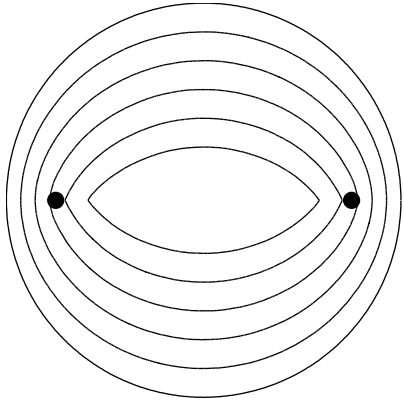
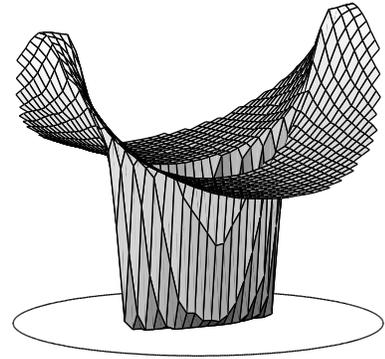
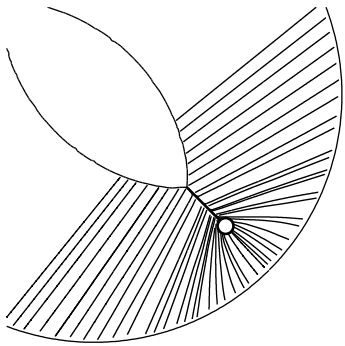
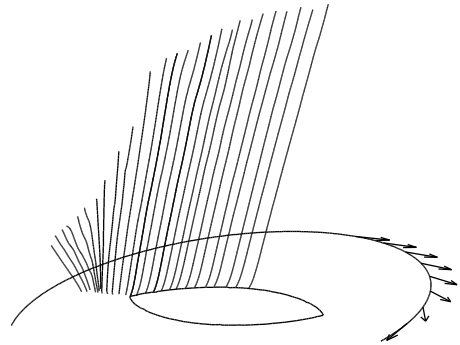
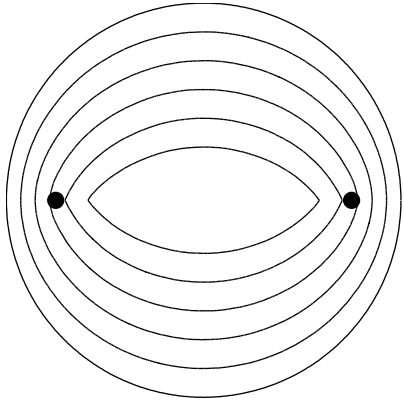
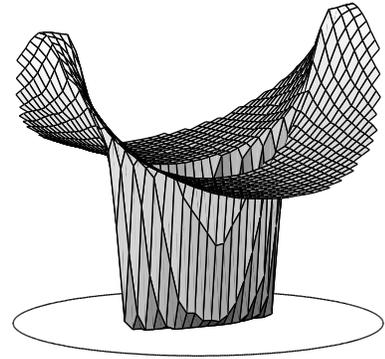
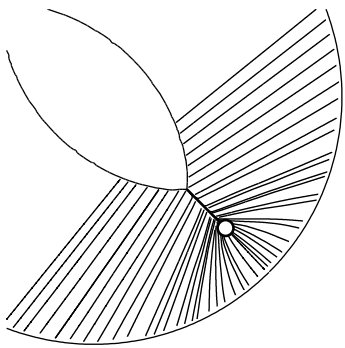
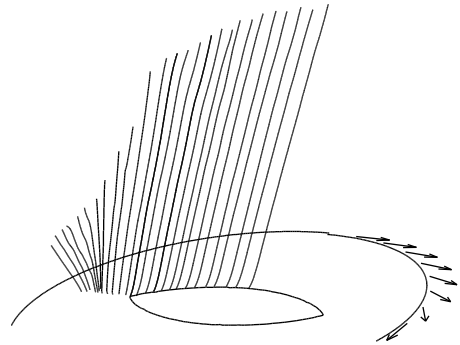
\end{document}